\documentclass{article}
\usepackage{booktabs}
\usepackage{float}
\usepackage{authblk}
\usepackage{amsmath}
\usepackage[letterpaper,top=2cm,bottom=2cm,left=3cm,right=3cm,marginparwidth=1.75cm]{geometry}
\usepackage[utf8]{inputenc}
\DeclareUnicodeCharacter{03BC}{$\mu$}
\DeclareUnicodeCharacter{2009}{$\,$}
\DeclareUnicodeCharacter{200A}{ }
\DeclareUnicodeCharacter{2212}{-}
\usepackage[backend=bibtex,style=numeric,sorting=none]{biblatex}
\usepackage{chemformula}
\usepackage{lineno}
\usepackage{nameref}
\usepackage{xurl}

\title{Coherent photo-thermal noise cancellation in a dual-wavelength optical cavity for narrow-linewidth laser frequency stabilisation}

\author[1,2,*]{Fabian Dawel}
\author[1,2]{Alexander Wilzewski}
\author[1]{Sofia Herbers}
\author[1]{Lennart Pelzer}
\author[1,2]{Johannes Kramer}
\author[1,2]{Marek B. Hild}
\author[1,2]{Kai Dietze}
\author[1,2]{Ludwig Krinner}
\author[1]{Nicolas C. H. Spethmann}
\author[1,2]{Piet O. Schmidt}

\affil[1]{\centering  Physikalisch-Technische Bundesanstalt, Bundesallee 100, 38116 Braunschweig, Germany}
\affil[2]{ Institut für Quantenoptik, Leibniz Universität Hannover, Welfengarten 1, 30167 Hannover, Germany}
\affil[*]{Corresponding author: Fabian.Dawel@ptb.de} 

\date{}

\begin{document}

\maketitle

\begin{abstract}
Optical resonators are used for the realisation of ultra-stable frequency lasers. The use of high reflectivity multi-band coatings allows the frequency locking of several lasers of different wavelengths to a single cavity. While the noise processes for single wavelength cavities are well known, the correlation caused by multi-stack coatings has as yet not been analysed experimentally. In our work, we stabilise the frequency of a $729\,$nm and a $1069\,$nm laser to one mirror pair and determine the residual-amplitude modulation (RAM) and photo-thermal noise (PTN). We find correlations in PTN between the two lasers and observe coherent cancellation of PTN for the $1069\,$nm coating. We show that the fractional frequency instability of the $729\,$nm laser is limited by RAM at $1\times10^{-14}$. The instability of the $1069\,$nm laser is at $3\times10^{-15}$ close to the thermal noise limit of $1.5\times10^{-15}$.
\end{abstract}

\section{Introduction}

Optical Fabry-Pérot cavities are an indispensable tool in precision experiments like gravitational wave detectors \cite{weiss_nobel_2018, thorne_nobel_2018, barish_nobel_2018}, opto-mechanical systems \cite{poot_mechanical_2012, mcclelland_advanced_2011, aspelmeyer_cavity_2014}, micro-cavities \cite{panuski_fundamental_2020, qiu_dissipative_2022}, or optical atomic clocks \cite{ludlow_optical_2015}. Nowadays, the best optical cavities used for optical clocks reach a fractional length instability of $4\times 10^{-17}$ at $1\,$s \cite{matei_1.5_2017}. This instability is limited by the thermal noise of the mirror coating. It has been proposed to reduce coating thermal noise by employing different materials and coating stacks \cite{yam_multimaterial_2015, steinlechner_thermal_2015, kimble_optical_2008, hong_brownian_2013}, and this was experimentally demonstrated by introducing aSi into a highly reflective \ch{SiO2}/\ch{Ta2O5} coating \cite{tait_demonstration_2020}.

Apart from laboratory experiments in well-controlled environments, a rising number of transportable optical clocks \cite{hannig_towards_2019, koller_transportable_2017,ohmae_transportable_2021,poli_transportable_2014, zeng_toward_2023, cao_compact_2017, origlia_towards_2018, grotti_geodesy_2018} also have clock laser stabilisation cavities with instabilities below $1.6\times10^{-16}$ at $1\,$s \cite{herbers_transportable_2022}. Additionally, there are cavity designs for space applications \cite{pitkin_gravitational_2011,hafner_ultra-stabile_2015, argence_prototype_2012, sanjuan_long-term_2019} requiring even smaller, lighter and more robust setups. 

A growing number of spectroscopy experiments use multiple atomic species. The second species is used for sympathetic sideband cooling on a narrow optical transition \cite{bruzewicz_dual-species_2019, guggemos_sympathetic_2015, cui_sympathetic_2018, tanaka_sideband_2015, rugango_sympathetic_2015} or readout via quantum logic \cite{schmidt_spectroscopy_2005, hannig_towards_2019, lin_quantum_2020, guggemos_frequency_2019, chao_observation_2019}. Therefore, the demand for stabilising multiple narrow-linewidth lasers by means of an optical cavity is increasing. Separate resonators for each wavelength are expensive and increase the footprint of the experiment, which is a challenge for various applications, such as transportable setups or space applications. One solution is to use highly reflective dual wavelength coating resonators.

Locking multiple lasers to one cavity with one mirror pair has been demonstrated in the past. Different groups locked multiple lasers for Doppler cooling and repumping on the same mirror pair, to reduce the experimental footprint \cite{milani_multiple_2017, wang_integrated_2020} and to compensate for thermal drifts  \cite{hill_dual-axis_2021}.  Multi-wavelength coatings are also often used in transfer-cavity approaches \cite{rohde_diode_2010, leopold_tunable_2016, subhankar_microcontroller_2019, yin_narrow-linewidth_2015, dai_linewidth-narrowed_2014, albrecht_laser_2012, uetake_frequency_2009, bohlouli-zanjani_optical_2006, kruk_frequency_2005, rossi_long-term_2002, riedle_stabilization_1994}.

In this work we present to our knowledge for the first time the stabilisation of two lasers to a single narrow-linewidth cavity, using a dual-wavelength coating for $1069\,$nm and $729\,$nm and analyse their frequency instability. The photo-thermal effect \cite{an_optical_1997,de_rosa_experimental_2002,braginsky_thermo-refractive_2000, black_enhanced_2004,rosa_experimental_2006, farsi_photothermal_2012, herbers_transportable_2022,fejer_thermoelastic_2004} was measured in order to estimate the resulting frequency instability caused by optical power fluctuations on the mirrors. In addition, the residual-amplitude modulation (RAM)\cite{whittaker_residual_1985, wong_servo_1985, li_measurement_2012, shen_systematic_2015} in the setup was characterised. In the present analysis, we focus on the correlation effects caused by the cavity on both lasers. For comparison, we performed measurements of the individual lasers with the other physically blocked. 
Coherent cancellation effects of the thermal noise between the different coating stacks were observed and characterised.

\section{Experimental setup}

In this experiment we used a $50\,$mm long, cubic, ultra-low expansion (ULE$^{\text{\textregistered}}$) glass spacer \cite{webster_force-insensitive_2011}. The mirror substrates with 0.5{"} diameter were made from ULE$^{\text{\textregistered}}$ glass. One mirror is flat while the other has a radius of curvature of $R_2=500\,$mm. The non-reflective side of each mirror is wedged ($0.5\,$\textdegree) to prevent etalons. The quarter-wave stacks of the coating for the $1069\,$nm light are on the substrate, with the $729\,$nm coating stack on top. The coating order was chosen such as to minimise the absorption losses in the coatings and maximise the transmission of the $729\,$nm light. High transmission is important for (self-)injection-locking a slave diode to the cavity-filtered light in order to improve spectral purity \cite{akerman_universal_2015, labaziewicz_compact_2007, hald_efficient_2005, nazarova_low-frequency-noise_2008, krinner_low_2023}. The coating stacks\footnote{Made by Layertec} were made of \ch{SiO2} and \ch{Nb2O5}. The finesse was measured via a ring-down measurement. This resulted in a finesse of $67.2\,$k and $50.1\,$k for the $1069\,$nm and $729\,$nm cavities, respectively. For this cavity the main contribution to the thermal noise limit is given by the mirror substrate, where the main limitation for times between $0.01\,$s to $10\,$s is from Brownian noise. The calculated instability flicker frequency floor for the $729\,$nm frequency instability was mod$\,\sigma_y=1.5\times10^{-15}$ using the method shown in \cite{numata_thermal-noise_2004}. The thermal noise limit for the $1069\,$nm cavity is around $20\,$\% lower than the $729\,$nm thermal noise floor, when only  the $1069\,$nm coating is considered. The difference in coating thickness has only a small impact since it is not the main contributor to the thermal noise limit. The power spectral density of the substrate's thermal noise of the $1069\,$nm cavity is $2$ ($3$) times larger than the coating thermal noise when the whole (only the $1069\,$nm) coating stack contributes to the Brownian noise. In the plots presented below, we always show the fractional frequency fluctuations of the $729\,$nm thermal noise limit.

The estimated frequency shift from acceleration is below the thermal-noise limit for typical laboratory environments. This is achieved by mounting the  spacer via four edges of the cube to obtain high mounting symmetry \cite{webster_force-insensitive_2011}. The measured sensitivity to accelerations at $10\,$Hz acceleration modulation is: $\sigma_x = 2.8(6)\times10^{-11}/\text{g}$, $\sigma_y = 6.9(3)\times10^{-11}/\text{g}$ and $\sigma_z = 7.7(5)\times10^{-11}/\text{g}$. In principle, insensitivities on the order of $10^{-12}/\text{g}$ should be feasible with this spacer for two of the three directions, as shown in previous work \cite{webster_force-insensitive_2011}.
For additional vibrational decoupling of the cavity, it was placed on an active vibration isolation platform\footnote{Tablestable, AVI600-M}. With the given sensitivity, we reach a fractional frequency instability contribution from vibrations of $<4\times10^{-16}$ at 1s.

To reduce temperature drifts, the resonator was thermally isolated. Two heat shields provide passive temperature stability and the outer heat shield was actively regulated by four peltier elements. The mounting and the heat shields are shown in \cite{hafner_ultra-stabile_2015}. The zero crossing of the resonator's coefficient of thermal expansion  was measured to be at $28.5(1.0)\,$°C. 
A 1\% fluctuation at a pressure of $2\times10^{-6}\,$Pa at the ion pump, will cause a frequency instability of $<1\times10^{-16}$. 

\begin{figure}[ht]
\centering
\includegraphics[width=0.8\textwidth]{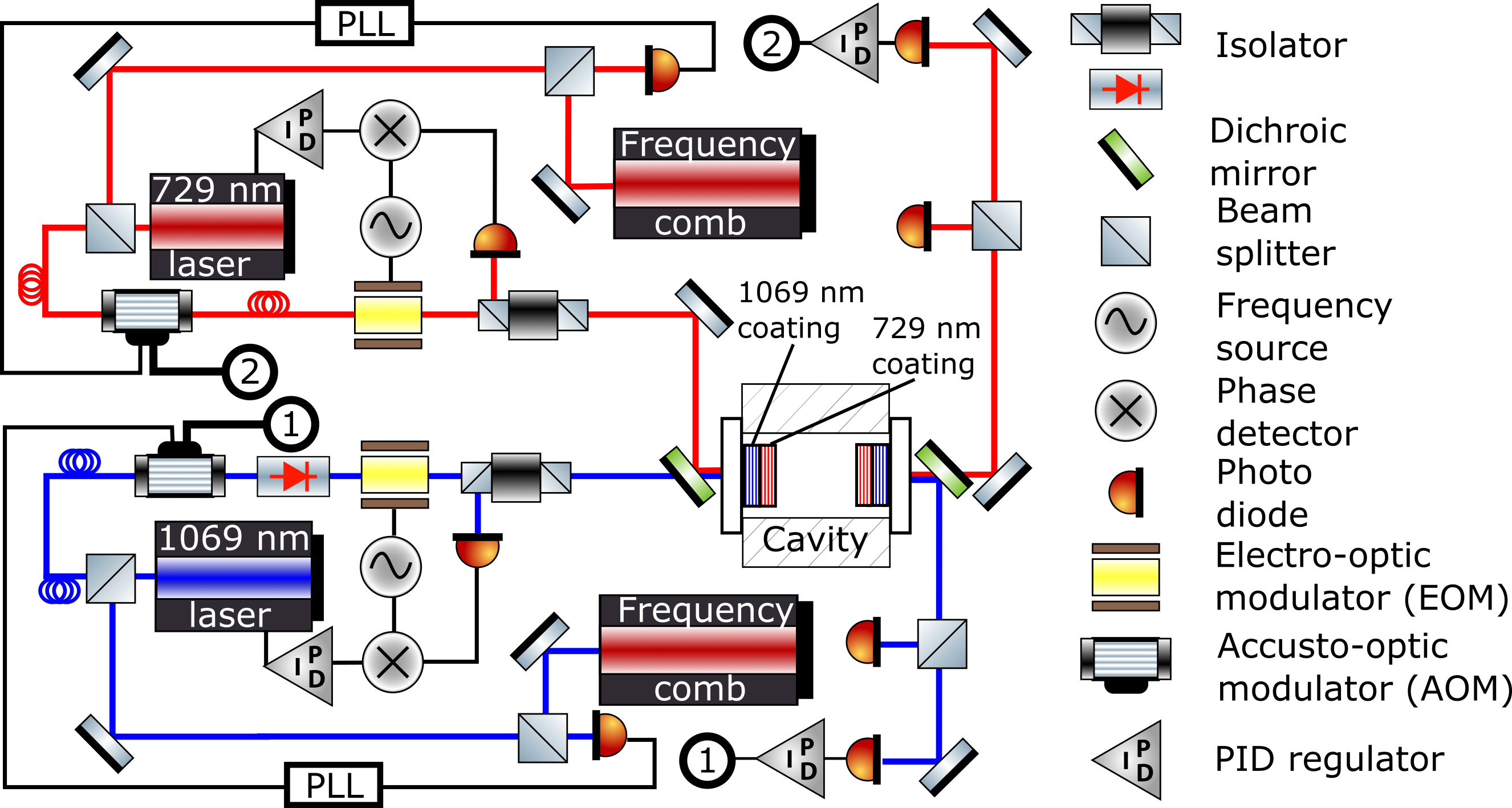}
\caption{Schematic drawing of the experimental setup. The red and the blue lines show the $729\,$nm laser path and $1069\,$nm laser path, respectively. Black lines represent electronic connections. PLL=Phase-locked loop; PID=Proportional-Integral-Derivative regulator \cite{ComponentLib}}
\label{fig:setup}
\end{figure}

The $1069\,$nm and the $729\,$nm lasers are both extended cavity diode lasers\footnote{Toptica DLpro}. To transfer the cavity's fractional length stability to the laser's fractional frequency stability, we used a Pound-Drever-Hall (PDH) locking scheme \cite{black_introduction_2001, pound_electronic_1946, drever_laser_1983}. The modulation frequency for the $1069\,$nm laser was $12.3\,$MHz, applied with a wedged electro-optical modulator (EOM)\footnote{Qubig, PM7-NIR+w}. The $729\,$nm laser was modulated at $12.7\,$MHz with a Brewster-cut EOM\footnote{Qubig, PM7-NIR+BC}. We used two branches for feedback via a PID regulator\footnote{Toptica, FALC}: the fast signal controlled the laser diode current and the slow signal the external cavity grating angle. The optical setup is depicted in figure \ref{fig:setup}. Both laser beams are superposed in front of the cavity with a dichroic mirror. The light transmitted through the cavity was separated with a dichroic mirror to stabilise the transmitted optical power of each laser individually. For each wavelength, the transmitted light beam was split and two photodiodes were used to separately detect power fluctuations. One photodiode signal was used for the control circuit and the second allowed for an out-of-loop measurement.

For the frequency instability measurement, both laser frequencies were compared to a stable reference laser via an optical frequency comb. The comb was locked to a $1542\,$nm laser with a fractional frequency instability of $4\times10^{-17}$ at $1\,$s \cite{matei_1.5_2017}. Applying a transfer beat \cite{stenger_ultraprecise_2002, scharnhorst_high-bandwidth_2015} we measured the instability of both lasers using a synchronous frequency counter in lambda counting mode \cite{kramer_multi-channel_2001}. By transfer locking both lasers to the frequency comb \cite{scharnhorst_high-bandwidth_2015} via a phase-locked loop, we were able to extract frequency and phase shifts from the error signal.

\section{Measurement and Results}

\subsection{Photo-thermal effect}
\label{sec:PTN}
Changes in the cavity resonance frequency can be caused by light power fluctuations. This is known as the photo-thermal effect \cite{an_optical_1997,de_rosa_experimental_2002, black_enhanced_2004, rosa_experimental_2006, farsi_photothermal_2012,herbers_transportable_2022}. The absorption of light increases the temperature in coating and substrate, causing thermal expansion. This process results in photo-thermal elastic (PTE) noise. Temperature changes can also affect the refractive index of the coating, altering the  optical-path length and thus the resonance condition \cite{braginsky_thermo-refractive_2000}. This process results in photo-thermal refractive (PTR) noise. In a dual-wavelength setup, both light beams are partially absorbed in the mirror coating. Thus, power fluctuations of both affect the cavity's resonances. We assume that the modulation of laser light power is slower than the heat exchange between all coating layers \cite{fejer_thermoelastic_2004}. This means that the coating layer temperature along the beam propagation direction is time independent, and that the power absorption of both lasers will cause the same optical length change for equal amounts of dissipated power.

\begin{table}[ht]
\centering
\begin{tabular}{l|cccccc}
      & Nb$_2$O$_5$   & SiO$_2$    & ULE    \\
\midrule
$\eta$ & 0.2     & 0.17   & 0.17   \\
$\alpha$(1/K) & $5.8\times10^{-6}$  & $2.2\times10^{-6}$ & $0\pm3\times10^{-8}$   \\
$\kappa$ (W/[Km])& assumed: 1       & 1.38   & 1.38   \\
Y (GPA)    & 68    & 72   & 67.6 \\
C (MJ/[m$^3$K])    & 2.71   & 1.64 & 1.7 \\
$\Phi$   & $4.6\times10^{-4}$  & $2\times10^{-4}$   & $1.6\times10^{-5}$ \\
$\beta$ (1/K)  & $1.43\times10^{-5}$ & $8\times10^{-6}$   &        \\
n ($1069\,$nm)     & 2.25    & 1.46   &      \\ 
n ($729\,$nm)     & 2.29    & 1.46   &      
\end{tabular}
\caption{Material parameters used. Here, $\eta$ is Poisson's ratio, $\alpha$ is the coefficient of thermal expansion, $\kappa$ is the thermal conductivity, Y is Young's modulus, C is the specific heat per volume, $\Phi$ is the mechanical loss factor, and $\beta$ the thermorefractive index. The parameters are taken from \cite{herbers_transportable_2022,franc_mirror_2009,refractiveindex} and references therein.}
\label{tab:values}
\end{table}

The theory of photo-thermal noise (PTN) for Bragg mirrors was first described in \cite{braginsky_thermodynamical_1999}. Here, we use the equations given in the \nameref{sec:Appendix}. Based on the material parameters (see Tab.\ref{tab:values}), the optical length change X caused by light power fluctuations can be calculated for a given mirror. The total length change X is caused by the PTR effect of the mirror coating $X^{(\mathrm{ct})}_{\mathrm{PTR}}$, the PTE effect of the mirror coating $X^{\mathrm{(ct)}}_{\mathrm{PTE}}$, and the PTE effect of the mirror substrate $X^{(\mathrm{sb})}_{\mathrm{PTE}}$. The total length change $X$ is a complex number whose absolute value $|X|$ can be interpreted as the mirror displacement and $arg(X)$ is the phase between the absorbed power change and the length change. Thus, a change of phase by $180\,$° changes the sign of X. The overall shift is then calculated by adding all three contributions:
\begin{equation}
     |X| = |X^{(\mathrm{ct})}_{\mathrm{PTR}} + X^{(\mathrm{ct})}_{\mathrm{PTE}} + X^{(\mathrm{sb})}_{\mathrm{PTE}}|.
\end{equation}
Since the PTE effect and PTR effect may be out of phase, it is possible to have a coherent cancellation between the three parts. The formulas can be used to design a coating with coherent cancellation between the PTR and PTE effect as shown in \cite{chalermsongsak_coherent_2016} to reduce PTN for a certain frequency range of the power noise spectrum.

The cavity's length sensitivity to power fluctuations can be measured by modulating the power of the laser light and hence the absorbed power. For the optical power calibration, we measured the transmitted laser light on a photodiode. We further expect that the measured power transmission through the cavity is proportional to the power circulating in the cavity and therefore to the absorbed power in the coatings. As the ratio between scattered and absorbed light is unknown, we cannot infer the absorbed power (see below). The amplitude and the phase of the laser frequency change can be extracted from the error signal of the transfer lock to the reference laser and were converted into an optical length change. This measurement was performed for power modulation frequencies from $10\,$mHz to $100\,$Hz.
We investigated two cases. In the first, each laser frequency was locked to the resonator resonance frequency while the other laser was physically blocked. For the second, we locked both lasers to the resonator's respective resonance frequencies and one laser was power modulated, and then vice versa. 

\begin{figure}[ht]
\centering
\includegraphics[width=0.9\linewidth]{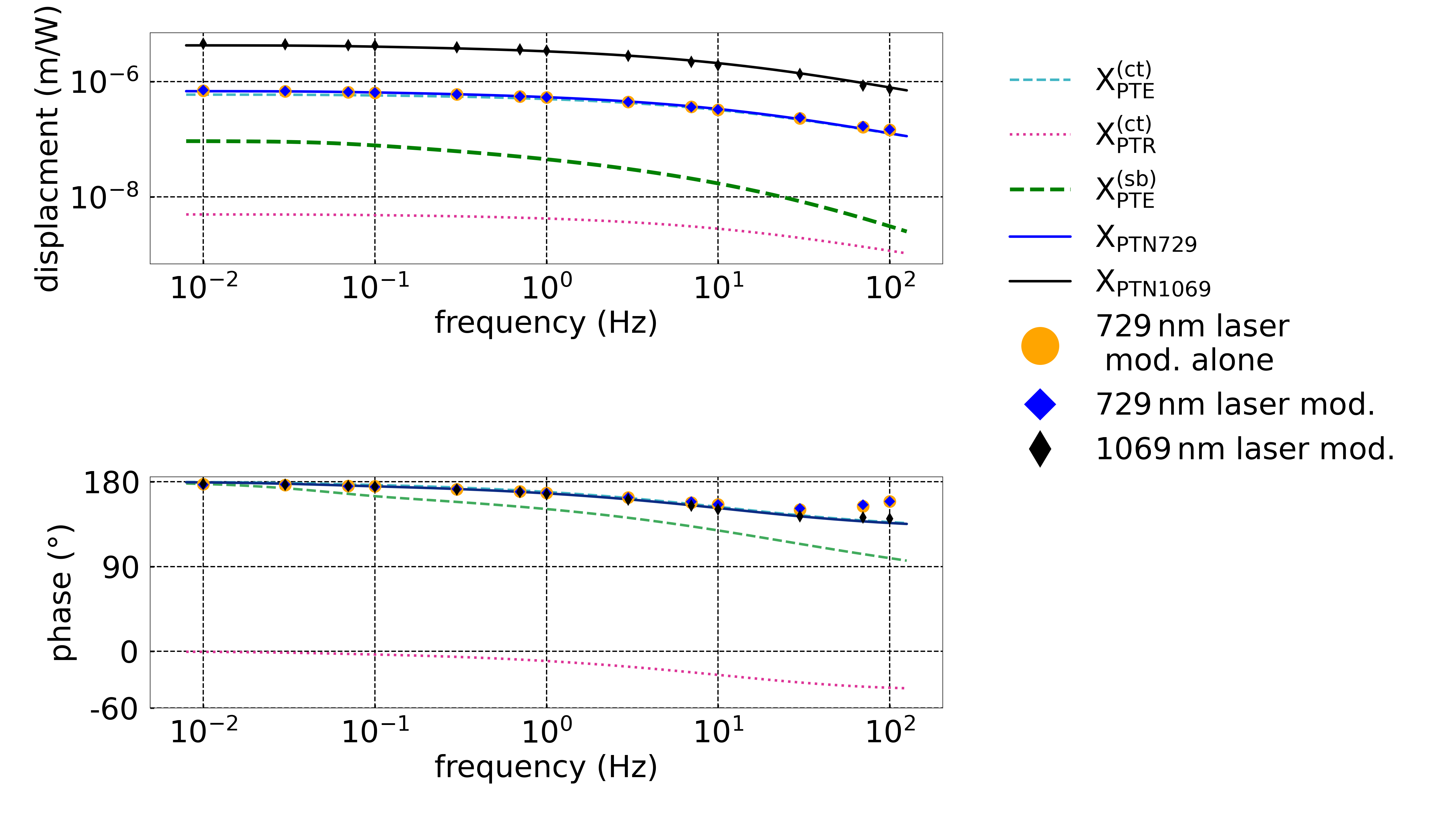}
\caption{Evaluation of the $729\,$nm laser frequency PTN sensitivity with theory fit to the measurement data. The upper figure shows the sensitivity and the lower one shows the phase relation. The laser modulated in power during the measurement is indicated by "mod.". The "alone" indicates that the $1069\,$nm laser was blocked. The shown measurement points are the average of three different power modulation settings. 
The contributions to the overall $729\,$nm PTN curves are shown for the $729\,$nm power modulation case with fitted material parameters. The difference in measured displacement between both lasers is due to the normalisation per transmitted power.}
\label{fig:PTN_729}
\end{figure}

Let us first look at the frequency response of the $729\,$nm laser frequency as a function of its light power modulation. Fig. \ref{fig:PTN_729} shows the corresponding optical length change of the resonator per transmitted power. The $729\,$nm laser power modulation showed no measurable difference between $1069\,$nm light blocked or frequency coupled to the resonator. The shown curves are fits of material parameters to account for effects such as the dependence on coating layer deposition\cite{braginsky_measurements_2003} or the use of substrate \cite{cetinorgu_mechanical_2009}.
We therefore use combination of material parameters as free parameters, i.e. $P_{\text{abs/trans}}\times\alpha_{\text{ct}}$, $P_{\text{abs/trans}}\times\beta_{\text{ct}}$, $C_{\text{ct}}$, $\kappa_{\text{ct}}$ and $P_{\text{abs/trans}}\times\alpha_{\text{sb}}$. Here, $P_{\text{abs/trans}}$ is the unknown scaling factor between transmission and absorption. As this is a fully correlated parameter, it cannot be determined by a fit independent from the material parameters. We fitted the data to the $1069\,$nm PTN sensitivity and $729\,$nm PTN sensitivity simultaneously, because the material parameters are the same in both cases. The fit parameters are shown in Tab.$\,$\ref{tab:fit_values}. The fitted curves of the substrate show that it does not contribute much to the $729\,$nm PTN sensitivity due to its low coefficient of thermal expansion (CTE). The expansion of the coating is the dominant contribution to the $729\,$nm laser frequency change. Most of the $729\,$nm power is reflected in the first layers, so that the $1069\,$nm stack does not make a significant contribution to the PTR noise.

\begin{table}[H]
\centering
\begin{tabular}{l|c||l|c}
      &            Literature & &Fit \\
\midrule
$\alpha_{\mathrm{ct}}$(1/K) &  $3.6\times10^{-6}$  &$P_{\mathrm{abs}/\mathrm{trans}}\times\alpha_{\mathrm{ct}}$(1/K) &   $1.80(10)\times10^{-5}$ \\
$\alpha_{\mathrm{sb}}$ (1/K)   & $\pm3\times10^{-8}$  & $P_{\mathrm{abs}/\mathrm{trans}}\times\alpha_{\mathrm{sb}}$(1/K) & $1.2(4)\times10^{-7}$  \\
$\kappa_{\mathrm{ct}}$ (W/[Km]) &$1.2$  & $\kappa_{\mathrm{ct}}$ (W/[Km])&  $1.04(10)$\\
$C_{\mathrm{ct}}$ (MJ/[m$^3$K]) &$2.1$   &$C_{\mathrm{ct}}$ (MJ/[m$^3$K])&  $1.37(16)$\\
$\beta_{\mathrm{ct}}^{729}$ (1/K) & $3.2\times10^{-6} $ &$P_{\mathrm{abs}/\mathrm{trans}}\times\beta_{\mathrm{ct}}^{729}$(1/K) & not fitted  \\
$\beta_{\mathrm{ct}}^{1069}$ (1/K) &$1.54\times10^{-4} $&$P_{\mathrm{abs}/\mathrm{trans}}\times\beta_{\mathrm{ct}}^{1069}$(1/K) & $3.02(10)\times10^{-4}$ \\ 
\end{tabular}
\caption{Summary of expected material values from the literature and the material values we obtained from the fits in Fig. \ref{fig:PTN_729} and \ref{fig:PTN_1069}. The expected values were calculated using the values in Tab. \ref{tab:values} and the equations in the  \nameref{sec:Appendix}. $\beta_{\mathrm{ct}}^{729}$ was not fitted because the measurement is not sensitive to the thermorefractive noise of the $729\,$nm coating. 
 $\beta_{\mathrm{ct}}^{1069}$ and $\beta_{\mathrm{ct}}^{729}$ were calculated using the thin-film transfer matrix method described in \cite{born_principles_2013,fowles_introduction_1989,ogin_measurement_2012}.}
\label{tab:fit_values}
\end{table}

Next we measured the effect of a power modulation of the $1069\,$nm laser light on the frequency response of the $729\,$nm laser frequency sensitivity, which is also depicted in Fig. \ref{fig:PTN_729}. The length change caused by $1069\,$nm laser light is higher compared to the previous case because of the higher absorption. This absorption increase is due to $1069\,$nm light passing through the $729\,$nm coating layers, which absorb a certain portion of the light.  We assume that the heat transfer in the coating in axial direction is faster than the laser power modulation, and that the only difference in PTN sensitivity for the $729\,$nm laser frequency is due to the amount of absorption and transmission differences between both lasers. As we scale to transmitted power, we do not account for this constant difference due to absorption and transmission. Therefore, we multiplied a constant absorption-to-transmission-ratio factor of $\approx6.2$ to the theory prediction fit of the $729\,$nm laser. This factor was calculated by the mean of the ratio of the $1069\,$nm and $729\,$nm laser power modulation data from Fig \ref{fig:PTN_729}. By using this rescaling we can fit all $729\,$nm and $1069\,$nm results with the same fit parameters. The frequency and phase behaviour due to $1069\,$nm power modulation is similar to the $729\,$nm power modulation case. So the absorption of light in lower layers of the coating does not change the PTN for the $729\,$nm laser.

\begin{figure}[ht]
\centering
\includegraphics[width=0.9\linewidth]{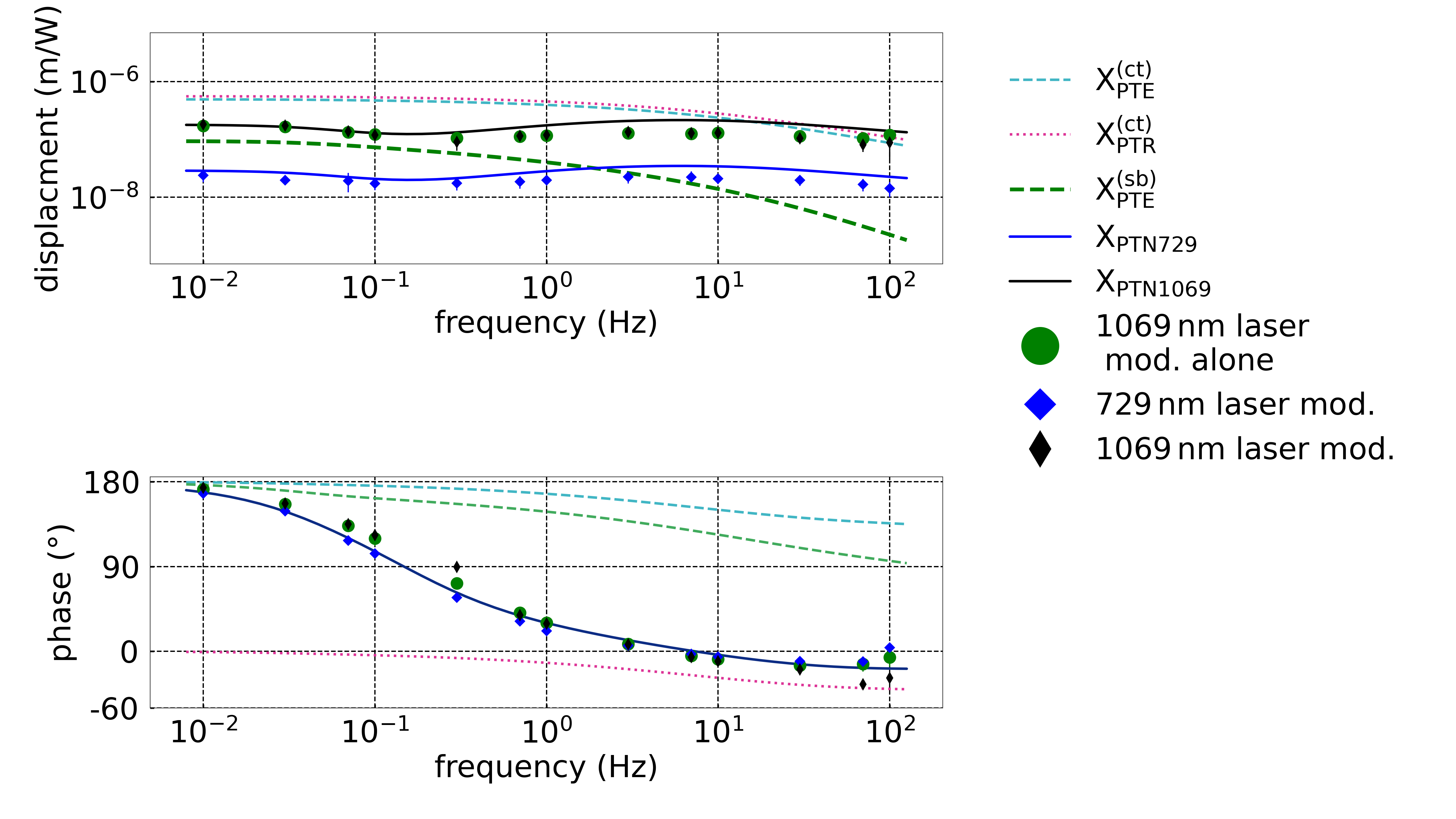}
\caption{Evaluation of the $1069\,$nm laser frequency PTN sensitivity with fit to measurement data. The upper figure shows the optical length change measured with the $1069\,$nm laser, while the lower figure shows the phase relation. The contributions to the overall $1069\,$nm displacement curves are shown for the $729\,$nm power modulation case with fitted material parameters. The shown measurement points are the average of three different power modulation settings.}
\label{fig:PTN_1069}
\end{figure}

The $1069\,$nm laser frequency sensitivity due to a modulation of optical laser power is shown in Fig. \ref{fig:PTN_1069}. Comparing Fig. \ref{fig:PTN_1069} to Fig. \ref{fig:PTN_729}, one can see that the $1069\,$nm PTN sensitivity is two orders of magnitude lower than the $729\,$nm PTN sensitivity. This is the case for power modulation by the $729\,$nm and $1069\,$nm laser. The PTE and PTR effect are out of phase and the amplitude of both effects is nearly identical. This leads to a coherent cancellation.
While the refractive index change of the $729\,$nm stack has an impact on the optical path length of the $1069\,$ nm light, the $1069\,$nm stack has no impact on the optical path length of the $729\,$nm light because the $1069\,$nm stack is positioned behind the $729\,$ nm stack. Thus, the PTR contribution to the $1069\,$nm laser's frequency (Fig. \ref{fig:PTN_1069}) change is much larger compared to that of the $729\,$nm laser (Fig. \ref{fig:PTN_729}) at the same light power modulation. The use of the transfer matrix approach implies that the amount of PTE noise is the same in both cases.

Using the fit results together with measurements of the absorption, loss and transmittance of the cavity, one can derive lower bounds for the material parameters. The loss of the $729\,$nm cavity is $39(1)\,$ppm and its transmission $23(1)\,$ppm measured in only one direction. Our measurements also showed that the $1069\,$nm cavity has a loss of $39(3)\,$ppm and a transmission of $11(3)\,$ppm. 
To obtain the lower boundaries of the material parameters, we assume that all losses for the $1069\,$nm laser light are absorption losses (no scattering loss) such that $P^{729}_{\mathrm{abs}/\mathrm{trans}}<0.59$, which gives us $\alpha_{\mathrm{ct}}^{\mathrm{meas}}>3.06(17)\times10^{-5}\,$1/K, $\alpha_{\mathrm{sb}}^{\mathrm{meas}}>2.0(7)\times10^{-7}\,$1/K, and $\beta_{\mathrm{ct}}^{1069,\mathrm{meas}}>5.11(17)\times10^{-5}$. 
For $\alpha_{\mathrm{ct}}$, the measured value is one order of magnitude larger than the literature value (see Tab. \ref{tab:fit_values}). This deviation can be contributed to a low $\alpha_{ct}^{\mathrm{SiO}_2}$ value. As discussed in \cite{ogin_measurement_2012,cetinorgu_mechanical_2009} the measured values between the bulk material or a thin film can differ because of microstructure, interface, and the underlying substrate. For the thermal expansion of Nb$_2$O$_5$, the reported literature values for $\alpha_{ct}^{\mathrm{Nb}_2\mathrm{O}_5}$ vary from $-2\times10^{-6}\,$1/K to $5.8\times10^{-6}\,$1/K \cite{manning_thermal_1972,choosuwan_negative_2002,cetinorgu_mechanical_2009}. The thermal refractive index $\beta_{\mathrm{ct}}^{1069}$ is also higher than the literature value (see Tab. \ref{tab:fit_values}) due to the larger CTE. The thermal expansion of the substrate  $\alpha_{sb}$ is higher than expected for ULE material. The mirrors and the spacer are of the same kind of ULE glass (not the same batch), so the mirrors are operated close to the thermal zero crossing. The large discrepancy is most likely due to the low sensitivity of the fit to the substrate's thermal expansion. The fitted material parameters indeed changed by no more than $2\sigma$ when $\alpha_{\mathrm{sb}}^{\mathrm{meas}}$ was set equal to the upper end of the literature value of $3.0\times10^{-8}\,$1/K. 
 The heat capacitance values exhibit a large spread as is known from the literature\cite{franc_mirror_2009}. Our value differs by at least $3\sigma$ compared to the closest literature value. For the thermal conductance, our value is in agreement with the literature value (see Tab. \ref{tab:fit_values}).

\begin{figure}[H]%
    \centering
    \includegraphics[width=0.9\linewidth]{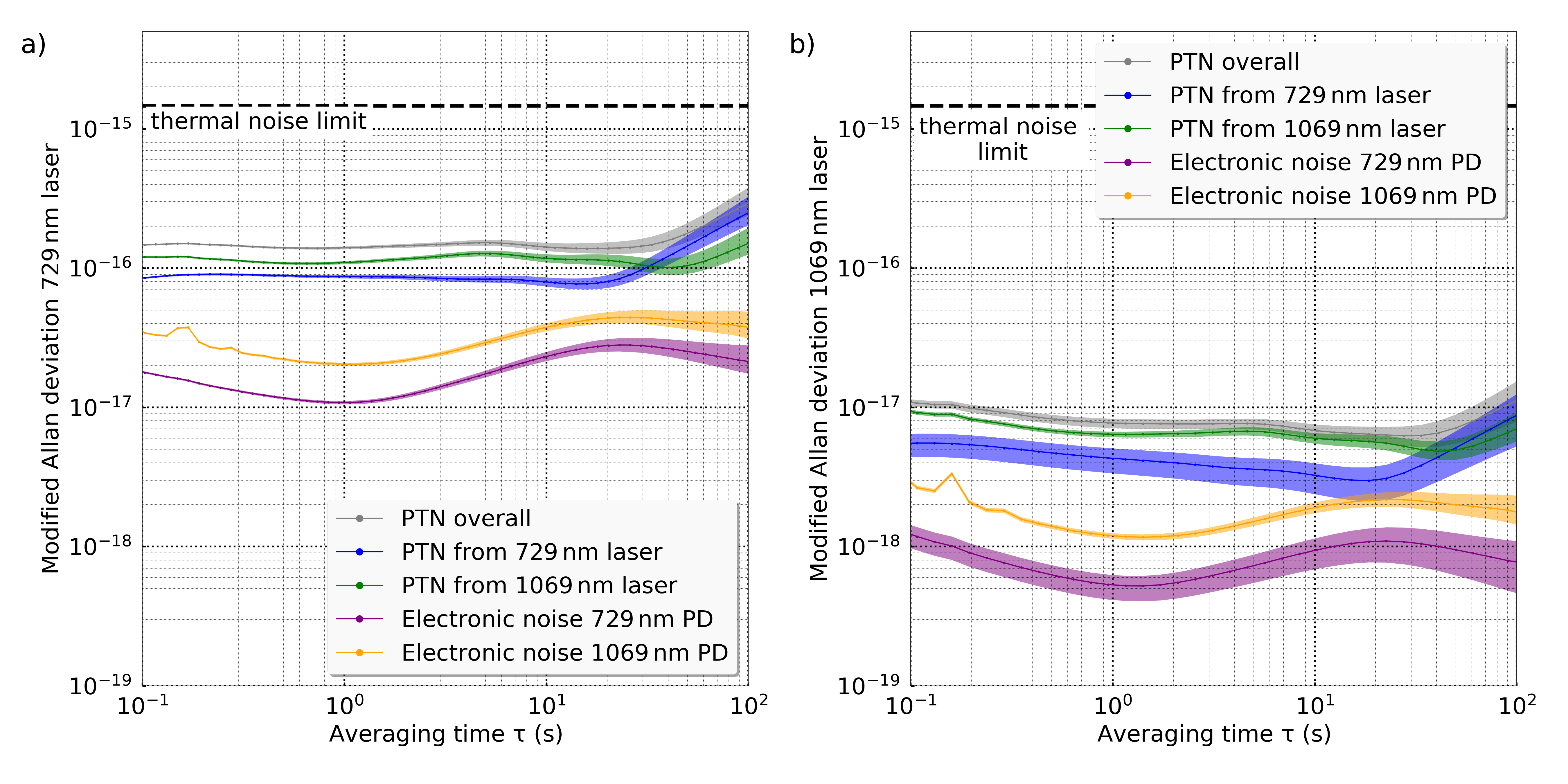}
    \caption{Evaluation of fractional frequency instability due to the PTN of each laser. Figure a) [b)] shows the PTN for the $729\,$nm [$1069\,$nm] laser. Both have a PTN below the thermal noise limit of the cavity. For reference, the electronic noise of the photodiodes is also shown.}%
    \label{fig:PTE}%
\end{figure}

With the PTN sensitivity determined in Fig. \ref{fig:PTN_729} and \ref{fig:PTN_1069}, we can estimate the laser frequency fluctuations due to laser power fluctuations in the resonator. Fig.\,\ref{fig:PTE} shows that the frequency instability for the $1069\,$nm and the $729\,$nm laser is not limited by PTN. For the measurement, both lasers are power stabilised using photodiodes behind the cavity. Two additional out-of-loop photodiodes measure the remaining power fluctuations of the light transmitted through the cavity.
The PTN of both cavities, $729\,$nm and $1069\,$nm, is mainly limited by the power fluctuations of the $1069\,$nm laser, which can be explained by the larger absorption-to-transmission ratio for the $1069\,$nm light compared to $729\,$nm light.

\subsection{Residual amplitude modulation (RAM)}
In the PDH lock the laser light needs to be phase modulated with a fixed frequency. The demodulation of the reflected light of the cavity converts laser frequency fluctuations into amplitude fluctuations of an electronic signal. Additional optical amplitude fluctuations at the modulation frequency are called RAM and result in an offset of the electronic error signal that can change with time and thus compromise the frequency stability of the laser. One cause of RAM is birefringence in the EOM \cite{  wong_servo_1985,li_measurement_2012}. Another cause is that parasitic etalons affect the light amplitude of the carrier and sidebands differently \cite{whittaker_residual_1985,shen_systematic_2015}. There are stabilisation schemes which can mitigate the effect of RAM \cite{shi_suppression_2018, dominguez_fundamental_2017, descampeaux_new_2021, shen_systematic_2015, zhang_reduction_2014, ishibashi_analysisreduction_2002, bi_suppressing_2019}. 

We investigated the fractional frequency instability added by RAM for two cases. First, we showed the fractional frequency instability caused by RAM for each laser by physically blocking the light of the other laser. Second, we stabilised both lasers to the cavity to observe cross-correlations which might increase the amount of RAM compared to the first case. 

We measured RAM using a method similar to the one described in  \cite{herbers_transportable_2022}. For this, we measured the laser frequency to voltage conversion (frequency discriminant) by modulating the PDH error signal with a known voltage and measured the corresponding frequency change at the frequency comb while the laser was locked to the cavity. This resulted in $\approx60\,\frac{\text{kHz}}{\text{V}}$ for both lasers. For the measurement of RAM we tuned the laser away from resonance (around $150\,$MHz). The voltage fluctuations were measured and multiplied by the frequency discriminant to get the frequency instability caused by RAM.

\begin{figure}[ht]%
    \centering
    \includegraphics[width=0.9\linewidth]{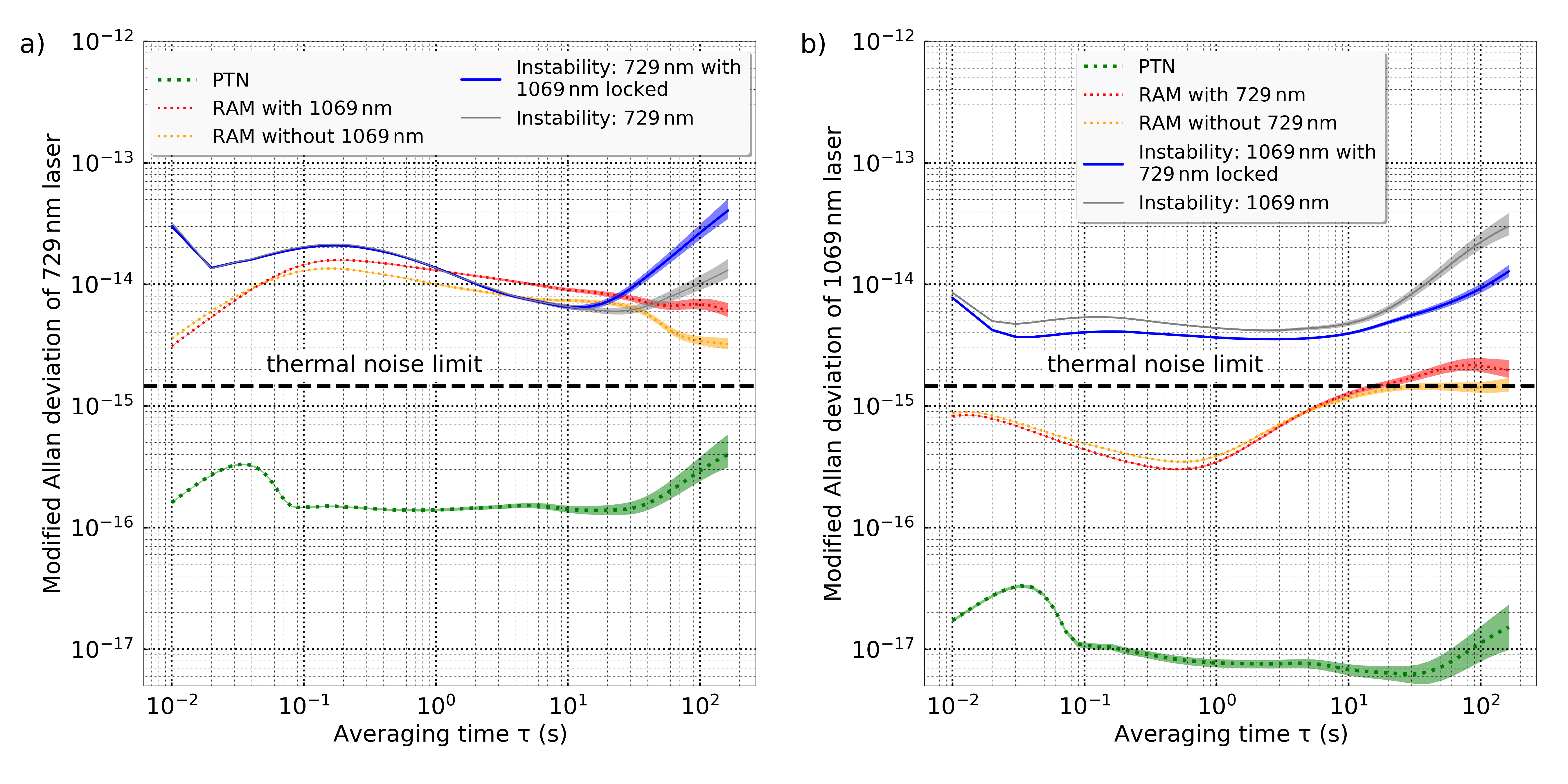}
    \caption{Evaluation of the laser instability. Figure a) shows the instability of the $729\,$nm laser including the effects of RAM and PTN. RAM clearly limits the frequency instability of the $729\,$nm laser. Figure b) shows the instability of the $1069\,$nm laser and the effects due to RAM and PTN. All shown modified Allan deviations for laser frequency instabilities are with linear drift subtracted. All errors shown in the RAM and lasers frequency instabilities only represent the statistical error of that measurement. They do not include larger errors due to day-to-day fluctuations or measurement device errors.}%
    \label{fig:instability}%
\end{figure}

 In Fig. \ref{fig:instability} a) [b)] we see the fractional frequency instability of the $729\,$nm [$1069\,$nm] laser caused by RAM. For the $729\,$ nm laser, the instability caused by RAM is above the thermal noise limit, therefore limiting the laser's fractional frequency instability. Typically, Brewster-cut and wedged EOMs introduce less RAM compared to other EOM types \cite{tai_electro-optic_2016,li_reduction_2016}. RAM voltage fluctuations of the order of $10\,$ppm (see, e.g., \cite{li_reduction_2016}) would decrease frequency fluctuations to a level of $10^{-15}$. Therefore, the birefringence of the EOM is unlikely to be the reason for the high RAM. We see indications for parasitic etalons due to a correlation between error signal and the scanned laser frequency far off-resonance. This was measured using a method similar to that in \cite{whittaker_residual_1985}. We found a modulation which could be caused by an etalon with a length of $5\,$m. This etalon is most likely located between the acousto-optical modulator and the optical isolator, therefore including the EOM (see Fig. \ref{fig:setup}). We do not expect that the RAM differs for the different cases since both lasers are phase modulated at different frequencies, which is corroborated by the near identical frequency instability of the $729\,$nm laser for the two cases. The slight difference in RAM for the two $729\,$nm laser cases can be most likely be attributed to temperature, pressure and humidity fluctuations in the lab over the course of the day and the limited measurement time for this long etalon. The frequency instability of the $1069\,$nm laser with the wedged EOM is not limited by RAM. The EOM shows RAM voltage fluctuations on a level of $1.6(0.1)\,$ppm (at $1\,$s). The instability caused by RAM is below the thermal noise limit for averaging times $<10\,$s. We observe a measurable difference in noise caused by RAM, whether the $729\,$nm laser is stabilised to the cavity, or blocked. As for the $729\,$nm laser, we do not find evidence for an influence of the second laser on the RAM of the $1069\,$nm laser.

\subsection{Fractional frequency instability}
The investigation of the laser fractional frequency instability is shown in Fig. \ref{fig:instability}. Here, the individually and simultaneously locked laser instabilities are shown. The instabilities for the $729\,$nm laser in both cases are nearly indistinguishable and they only separate for long averaging times $>10\,$s. The laser instability appears lower than the level expected by the amount of RAM measured (for $1-10\,$s). Typically, etalons causing RAM noise vary slowly over time for example due to changes in air temperature and other laboratory conditions.  Subsequent measurements can therefore yield slightly different RAM fluctuations. 

For the $1069\,$nm laser the the noise is very similar but the frequencies in the locked and unlocked cases can differ by a small amount. The difference between the measurements is limited by differential optical path length changes through optical fibres, or by small RAM fluctuations caused by the laboratory conditions changing during the experiments, and therefore differ between measurements.

\section{Conclusion}

We investigated the possibility of achieving narrow-linewidth lasers at very different optical frequencies by locking them to the same mirror pair. 
The laser instability due to RAM is not correlated between the two lasers. In the case of PTN, we could see a clear correlation between frequency shift and power modulation. The light absorption of both lasers contributes to the PTN of each laser. The effect of PTN on the fractional frequency instability of the $729\,$nm laser is below $2\times10^{-16}$($1\,$s) and thus smaller than the thermal noise limit. For the fractional frequency instability of the $1069\,$nm laser, the PTN sensitivity was even below $1\times10^{-17}$($1\,$s) due to a coherent cancellation between PTR and PTE. The overall laser fractional frequency instability is $3\times10^{-15}$ for the $1069\,$nm laser, which is around two times larger than the estimated thermal noise limit. We attributed the discrepancy to non-stabilised path lengths in the optical setup. For the $729\,$nm frequency lock, the fractional frequency instability was limited by RAM caused by a parasitic etalon and it only reached a level of $1.3 \times 10^{-14}$($1\,$s). This can be improved by eliminating the etalon, by finding the reflective surface, or by suppressing it using additional optical isolators. In general the thermal-noise floor and the insensitivity to RAM can be further improved by employing longer cavity spacers, thereby being able to achieve even lower length instability. In conclusion, we  demonstrated that dual-wavelength coatings provide a performance similar to single-wavelength coatings in narrow-linewidth laser stabilisation and can even reduce thermal noise components through coherent cancellation.

\section*{Funding}
Funded by the Deutsche Forschungsgemeinschaft (DFG, German Research Foundation) under Germany’s Excellence Strategy – EXC-2123 QuantumFrontiers – 390837967 and Project-ID 274200144 – SFB 1227, project B03. This joint research project was financially supported by the State of Lower Saxony, Hanover, Germany, through Niedersächsisches Vorab. This project also received funding from the European Metrology Programme for Innovation and Research (EMPIR) cofinanced by the five participating  States and from the European Union’s Horizon 2020 research and innovation programme (Project No. 20FUN01 TSCAC). This project further received funding from the European Research Council (ERC) under the European Union’s Horizon 2020 research and innovation programme (grant agreement No 101019987).

\section*{Acknowledgements}
The authors would like to thank Thomas Legero for the optical contacting of the cavity mirrors. We would also like to thank Uwe Sterr and Benjamin Kraus for fruitful discussions.

\section*{Disclosures}
The authors declare no conflicts of interest.

\section*{Data Availability Statement}
Data underlying the results presented in this paper are not publicly available at this time but may be obtained from the authors upon reasonable request.

\section*{Appendix}
\label{sec:Appendix}
The equations for calculating the photo-thermal noise are given here solely for the sake of completeness and were taken directly from \cite{herbers_transportable_2022}
\begin{equation}
    \delta X_{\mathrm{PTE}}^{\mathrm{(sb)}}=-\frac{\alpha_\mathrm{sb}[1+\eta_\mathrm{sb}]}{\pi\kappa_\mathrm{sb}}\frac{f_\mathrm{T}^\mathrm{(sb)}}{if}\delta P_{\mathrm{abs}}\int_0^\infty\zeta\mathrm{exp}\left(-\frac{\zeta ^2}{2}\right)F\left[1-\frac{\zeta }{\zeta_{sb}}\right]\,\mathrm{d}\zeta 
\end{equation}

\begin{equation}
    \begin{aligned}
        \delta X_{\mathrm{PTE}}^{\mathrm{(ct)}} = &-\frac{\alpha_\mathrm{ct}}{\pi\kappa_\mathrm{ct}}\frac{f_\mathrm{T}^\mathrm{(sb)}}{if}\delta P_{\mathrm{abs}}\int_0^\infty\zeta\mathrm{exp}\left(-\frac{\zeta ^2}{2}\right)F \\ &\times \biggl\{ \gamma_1\left[\cosh\left(\zeta d_{\mathrm{ct}}\sqrt{2}/\omega\right)+ R\frac{\zeta}{\zeta_{ct}}\sinh\left(\zeta d_{\mathrm{ct}}\sqrt{2}/\omega\right) \right.\\  &\left. -\cosh\left(\zeta_\mathrm{ct} d_{\mathrm{ct}}\sqrt{2}/\omega\right) - R\sinh\left(\zeta_\mathrm{ct} d_{\mathrm{ct}}\sqrt{2}/\omega\right)\right] \\ &-\gamma_2\frac{\zeta}{\zeta_\mathrm{ct}}\left[ R\cosh\left(\zeta d_{\mathrm{ct}}\sqrt{2}/\omega\right)+\frac{\zeta_\mathrm{ct}}{\zeta}\sinh\left(\zeta d_{\mathrm{ct}}\sqrt{2}/\omega\right) \right. \\ & \left. -R\cosh\left(\zeta_\mathrm{ct} d_{\mathrm{ct}}\sqrt{2}/\omega\right)-R\sinh\left(\zeta_\mathrm{ct} d_{\mathrm{ct}}\sqrt{2}/\omega\right)\right] \biggr\}
        \mathrm{d}\zeta 
    \end{aligned}
\end{equation}

\begin{equation}
    \begin{aligned}
        \delta X_{\mathrm{PTR}}^{\mathrm{(ct)}} = &\frac{\lambda_0\beta_{\mathrm{ct}}}{\sqrt{2}\pi\kappa_\mathrm{ct}\omega}\delta P_{\mathrm{abs}}\int_0^\infty\mathrm{exp}\left(-\frac{\zeta ^2}{2}\right)\frac{\zeta}{\zeta_\mathrm{ct}} \\ & \times  \frac{\sinh{\left(\zeta_\mathrm{ct} d_{\mathrm{ct}}\sqrt{2}/\omega\right)} +R\cosh{\left(\zeta_\mathrm{ct} d_{\mathrm{ct}}\sqrt{2}/\omega\right)}}{\cosh{\left(\zeta_\mathrm{ct} d_{\mathrm{ct}}\sqrt{2}/\omega\right)} +R\sinh{\left(\zeta_\mathrm{ct} d_{\mathrm{ct}}\sqrt{2}/\omega\right)}}
        \mathrm{d}\zeta 
    \end{aligned}
\end{equation}

Where the variables are defined as 
\begin{equation}
    F=\frac{1}{\cosh\left(\zeta_\mathrm{ct} d_{\mathrm{ct}}\sqrt{2}/\omega\right)+R\sinh\left(\zeta_\mathrm{ct} d_{\mathrm{ct}}\sqrt{2}/\omega\right)}
\end{equation}

\begin{equation}
    R=\frac{\kappa_\mathrm{ct}\zeta_\mathrm{ct}}{\kappa_\mathrm{sb}\zeta_\mathrm{sb}}
\end{equation}

\begin{equation}
    \zeta_\mathrm{ct}=\sqrt{i\frac{f}{f_\mathrm{T}^\mathrm{(ct)}}+\zeta^2}
\end{equation}

\begin{equation}
    \zeta_\mathrm{sb}=\sqrt{i\frac{f}{f_\mathrm{T}^\mathrm{(sb)}}+\zeta^2}
\end{equation}

\begin{equation}
    f_\mathrm{T}^\mathrm{(ct)}=\frac{\kappa_{ct}}{\pi\omega^2C_\mathrm{ct}}
\end{equation}

\begin{equation}
    f_\mathrm{T}^\mathrm{(sb)}=\frac{\kappa_{sb}}{\pi\omega^2C_\mathrm{sb}}
\end{equation}

\begin{equation}
    \gamma_1=\frac{1}{2}\frac{1-\eta_\mathrm{ct}}{1-\eta_\mathrm{ct}}\left[1+(1-2\eta_\mathrm{sb})\frac{1+\eta_\mathrm{sb}Y_\mathrm{ct}}{1+\eta_\mathrm{ct}Y_\mathrm{sb}}\right]
\end{equation}

\begin{equation}
    \gamma_2=\frac{1-\eta_\mathrm{sb}Y_\mathrm{ct}}{1-\eta_\mathrm{ct}Y_\mathrm{sb}}
\end{equation}
To get an average of the thermal expansion, heat capacitance and thermal conductance of the coating we weight the expansion by the thickness of each quarter-wave stack. Here we include the number $N_\mathrm{H}$ $(N_\mathrm{L})$ of high (low) refractive index stacks of the $1069\,$nm and $729\,$nm stacks.
\begin{equation}
    \alpha_\mathrm{ct}=\sum_{\mathrm{k}=1}^{N_L+N_H}\alpha_\mathrm{k}\frac{d_\mathrm{k}}{d_\mathrm{ct}}
\end{equation}

\begin{equation}
    C_\mathrm{ct}=\sum_{\mathrm{k}=1}^{N_L+N_H}C_\mathrm{k}\frac{d_\mathrm{k}}{d_\mathrm{ct}}
\end{equation}

\begin{equation}
    \kappa_\mathrm{ct}=\left(\sum_{\mathrm{k}=1}^{N_L+N_H}\frac{1}{\kappa_\mathrm{k}}\frac{d_\mathrm{k}}{d_\mathrm{ct}}\right)^{-1}
\end{equation}

\printbibliography

\end{document}